\documentclass[submission, Phys]{SciPost}
\usepackage{amsmath}
\usepackage{amsfonts}
\usepackage{graphicx}
\usepackage{float}
\usepackage{appendix}
\usepackage{dsfont}
\usepackage{amssymb}
\usepackage{bm}
\usepackage{color}
\usepackage{ulem}
\usepackage{soul}
\setstcolor{red}
\usepackage{nccmath}
\usepackage{array}

\newcommand{\beq}{\begin{equation}}
\newcommand{\eneq}{\end{equation}}
\newcommand{\be}{\begin{equation}}
\newcommand{\ee}{\end{equation}}
\newcommand{\bea}{\begin{eqnarray}}
\newcommand{\eea}{\end{eqnarray}}

\usepackage{multirow}
\usepackage{wasysym}

\begin{document}

\begin{center}{\Large \textbf{
Traffic models and traffic-jam transition in quantum ($N$+1)-level systems
}}\end{center}

\begin{center}
A. Nava\textsuperscript{*},
D. Giuliano\textsuperscript{},
A. Papa\textsuperscript{},
M. Rossi\textsuperscript{}
\end{center}

\begin{center}
{\bf} Dipartimento di Fisica, Universit\`a della Calabria, Arcavacata di Rende I-87036, Cosenza, Italy \\
INFN - Gruppo collegato di Cosenza,
Arcavacata di Rende I-87036, Cosenza, Italy
\\
* andrea.nava@fis.unical.it
\end{center}

\begin{center}
\today
\end{center}


\section*{Abstract}
{\bf
We propose a model to implement and simulate different traffic-flow conditions in terms of quantum graphs hosting an ($N$+1)-level
dot at each site, which allows us to  keep track of the type and of the destination of each vehicle.
By implementing proper Lindbladian local dissipators, we derive the master equations
that describe the traffic flow in our system.  To show the versatility and the reliability of our technique, we employ
it to model  different types of traffic flow (the symmetric three-way roundabout and the three-road intersection). Eventually, we
successfully  compare our predictions with results from classical models.
}

\vspace{10pt}
\noindent\rule{\textwidth}{1pt}
\tableofcontents\thispagestyle{fancy}
\noindent\rule{\textwidth}{1pt}
\vspace{10pt}

\section{Introduction}
\label{intro}

In the last decades the traffic behavior and the associated traffic-jam transition have been widely studied within the master equation
formalism for stochastic exclusion processes of many-body systems, using  models such as, for instance, the asymmetric
exclusion process (ASEP) as well as the total asymmetric exclusion process (TASEP)~\cite{Trimper, tasep_1, tasep_2, Sandow, Derrida}. All these models rely on the possibility to simulate the
dynamical behavior of the system at a coarse graining scale with vehicles living on a lattice with many-body interactions, rather than considering
a continuum of possible states (representing vehicles position and velocity). Therefore, the evolution of the system is not expressed in terms of
deterministic equations of motion, but by means of a set of stochastic rules that describe all the different paths that a system can follow starting
from a given initial state~\cite{ca_traffic_1,ca_traffic_2}. In fact, despite the traffic flow is a classic problem and, therefore, it
might seem appropriate to describe it by resorting to appropriate  hydrodynamic models,   the finite dimensions of vehicles introduce a natural spatial
``quantization'' of space. Indeed, this concept has been widely used in
cellular-automaton descriptions of the traffic-flow problem,
where both space and time are discrete~\cite{ca_traffic_1,ca_traffic_2, nagatani_1}
(incidentally, it is worth stressing how a similar ``quantization'' of both space and time, followed by
a systematic implementation of cellular-automaton approach, has proved to be pretty effective in, {\it e.g.},
the analysis of the infection spread in real space~\cite{is_1,is_2}).

In Ref.~\cite{Doi},  the stochastic  rules have been encoded in the transition probabilities for elementary evolutions
from one state to another. Moreover, they have been shown to be equivalent to the consolidated cellular automaton approach of Biham and
Nagatani~\cite{nagatani_1, nagatani_2, nagatani_3}. Of particular relevance is the fact that the  rules can be converted into a
set of linear operators acting on the vectors of a proper Fock space, taking advantage of the  formal analogy between classical stochastic processes
and quantum mechanical formalism. The Fock space description has been developed in terms of both
Bosonic and Fermionic operators on periodic and open systems (see, for instance,  Ref.~\cite{schutz} for a review on the subject).

Within the Fock space formulation of the problem, one  implements the analogy with quantum mechanical systems by
introducing  an effective Hamiltonian  $\hat{H}$, as well as vehicles creation and annihilation operators, similar to creation and annihilation operators
in quantum mechanics. Yet, the expectation value of an observable $\hat{O}$ for a system in the state $\left| \psi \right\rangle$ has  not
to be computed as $\left\langle \psi \right| \hat{O} \left| \psi \right\rangle$, but with respect to the ``sum-vector''
$\left\langle s \right|=\sum \left\langle n \right|$ with ${ \left| n \right\rangle }$ being  a basis of the Fock space, {\it i.e.},
it is defined as  $\left\langle s \right| \hat{O} \left| \psi \right\rangle$.

Despite its potentially wide applicability, the formalism based on the effective Hamiltonian  $\hat{H}$ is not easy to
implement, since it is not easy to compute the operator expectation values.
For this reason in the last years a fully quantum formulation in terms of a Markovian Lindblad master equation description for density matrix
has been introduced~\cite{Temme,paletta}, which  allows for  expressing  the stochastic classical rules in terms of Lindblad jump operators that
ensure (vehicle) particle conservation exclusion processes by allowing, at the same time, for an easy implementation of open boundary conditions.
At the same time this approach gives full access to single operator expectation values as well as to local and nonlocal, both in space and time, equilibrium correlation functions that are fundamental quantities for the characterization of the traffic behavior and for traffic forecasting and control~\cite{forecast_1,forecast_2}.
Furthermore, the approach allows to treat at the same level classical incoherent evolution and pure quantum coherent evolution opening
 the possibility to use quantum dot systems as experimental devices to simulate classical
traffic behavior~\cite{Non-Hermitian, multilevel}.

Remarkably, while a huge amount of
work has been already done for traffic description in one-dimensional systems~\cite{class_line_1, pre_3, pre_4, pre_5, pre_10, pre_11},
in the last years a  wide interest is arising around the possibility to describe different kinds of
intersections~\cite{pre_2, pre_8, pre_9}, searching for optimization procedures~\cite{pre_1, pre_6, pre_7, pre_12}.

As pointed out in Refs.~\cite{uta_1,kerner}, freeway traffic and urban traffic are qualitatively different phenomena, the first being governed by the interactions between vehicles, and the second by the priority rules at intersections.
Along this line of research, in this paper we analyze   graphs with multiple in/out sites in terms
of multilevel quantum dots. In particular, we are interested in the intersection between three possible routes where the exit
direction of each vehicle is not stochastically chosen during the hopping process, but it is rather an internal property of the vehicle.

Specifically, we discuss in detail  three-road junctions, which we model in terms of a quantum system on a graph, realized by
means of multilevel quantum dots coupled to external  reservoirs. Within our formalism, we use quantum hopping operators between quantum dots
to represent the flux of vehicles in  the junction. At variance, to account for vehicles entering the junction from outside/exiting from
the junction, we introduce hopping operators from/to external reservoirs to/from the internal dots.
Making use of quantum Lindblad master equation formalism we recover results in agreement with classical experiments.
  In doing so, we develop and discuss a three-dot model, which encodes the general information about the intersection of three routes, and
eventually generalize it to a six-dot model. This  allows us to implement different priority rules such as  a roundabout,
a right-hand priority junction, and an intersection between a major and a minor road. Remarkably,
despite the ``minimal'' setup we employ, our model exhibits all the relevant features in space and time
of a classical   traffic-flow diagram, such for instance the existence of a critical
density of vehicles beyond which the  traffic-jam phase transition sets in~\cite{nagatani_1,nagatani_2,nagatani_3}.

Moving backwards along the correspondence with quantum lattice systems, our model
can be regarded as a ``minimal'', pertinently adapted, version of the Y-junction of one-dimensional fermionic~\cite{oca_1,oca_2,guerci,buccheri_1} and/or bosonic~\cite{boso_0,boso_1,boso_2} quantum systems, including spin chains~\cite{tsve_1,crampettoni,gsst,gns,Giuliano_Campagnano,Fioravanti_2005}, as well as junctions  hosting the remarkable ``topological'' realization
of Kondo effect~\cite{beri_1,beri_2,topo_4,spin_charge_kondo,buccheri_2}, which have been largely addressed in the recent literature on correlated
quantum systems as simplest examples of devices exhibiting nontrivial phases/quantum phase transitions in their phase diagram.

The paper is organized as follows:

\begin{itemize}
 \item In Section \ref{model} we present our multi-directional model, in which we represent vehicles  as hardcore bosons on a graph.
 Consistently with our identification, we define
 the creation and annihilation operators for vehicles and write the Markovian Lindblad equation governing the time evolution.

 \item In Section \ref{cross} we present a minimal three-dot model that allows us  to address the main features of the
 three-route intersection.

 \item In Section \ref{sixdot} we extend the analysis of Section \ref{cross} by introducing a six-dot model that allows
 us to implement  different priority rules.

 \item In Section \ref{conclusion} we summarize our results and discuss possible  perspectives of our work.

\end{itemize}

\section{The model}
\label{model}

We now provide our  lattice Fock space description of the traffic flow in terms of multilevel quantum dots on
an open quasi-one-dimensional, multiple way out network  by also accounting for the possibility of having different
vehicles types or driving styles. To do so,
we divide a generic road into sections of length $a$, where $a$ is the mean dimension of a vehicle. Therefore, we describe
each section as a quantum $(N+1)$-level dot. The state $\left|0\right\rangle $ corresponds to an empty road
section, while the other $N$ states describe different vehicle types and destinations.  Assuming $m$ different vehicle types and $n$
different destinations, we have $N=m\times n$ 'occupied' states, which we label as $|j\rangle =\left|\left(v-1\right)n+d\right\rangle $
with integers $v\in\left\{1,\ldots,m\right\}$ and $d\in\left\{1,\ldots,n\right\}$.
Within our notation, the levels $|j\rangle$, with $j=1,...,n$, represent a vehicle of type $1$ directed towards
one of the $n$ possible destinations, the levels $|j\rangle$, with $j = n+1, \ldots, 2n$ represent a vehicle of type $2$ directed towards
one of the $n$ possible destinations, and so on. On each site we
introduce the   operators: $\sigma_{j,0}=\left|j\right\rangle \left\langle 0\right|$
($\sigma_{0,j}=\left|0\right\rangle \left\langle j \right|$),
that create  (destroy) a given vehicle-destination combination,
and $\sigma_{j', j}=\left|j'\right\rangle \left\langle j \right|$, with $(v-1)n+1 \leq j,j' \leq vn$,
that describe a vehicle of type $v$ changing its destination. These operators satisfy the conditions
$\sigma_{i,j}\sigma_{k,\lambda}=\delta_{j,k}\sigma_{i,\lambda}$, $\sigma_{i,j}^\dagger=\sigma_{j,i}$
and $\sum\limits_{j=0}^N \sigma_{j,j}=1$. As a consequence the dot can only be empty or occupied by a
single vehicle/destination combination as it is impossible to create more than a vehicle on the same site.

We can now build up the Hilbert space for a complex network of $L$ multi-level dots as the tensor product
of the Hilbert spaces of each dot. It is more convenient, as we will show later, to let $N$
be site dependent. Therefore,  we have, in general, $\mathcal{D}=\prod\limits_{\ell=1}^{L}(N_{\ell}+1)$ basis
vectors in the enlarged space, $\left|j_{1},\ldots,j_{L}\right\rangle$.
They are defined as the Kronecker tensor product of the basis vectors of each dot

\begin{equation}
\left|j_{1},\ldots,j_{L}\right\rangle =\left|j_{1}\right\rangle \otimes\ldots\otimes\left|j_{L}\right\rangle \, .
\end{equation}
In the same way, we label as $\sigma^{\left(\ell\right)}_{i,j}$ the  creation, annihilation and conversion operators acting on a given dot $\ell$:
these are realized as  the tensor product of ($L-1$)  identity matrices of dimension $N_{\ell} \times N_{\ell}$ and the $\sigma_{i,j}$ operator at the site $\ell$,
that is

\begin{equation}
\sigma_{i,j}^{\left(\ell\right)}=\mathbb{I}\otimes\ldots\otimes\sigma_{i,j}\otimes\ldots\otimes\mathbb{I} \, .
\end{equation}
For example, the state $\left|1,0,2\right\rangle$ represents a graph of three dots where we have an empty road on site two,
while sites one and three are filled by two different vehicles (or same vehicle type with different destinations); the operator
$\sigma_{0,j}^{\left(1\right)}=\sigma_{0,j}\otimes\mathbb{I}\otimes\mathbb{I}$ destroys the vehicle $j$
on site 1, and so on.

Having built an appropriate Fock space with intrinsic exclusion and multi-directional vehicles, we only need to introduce the
master equations that determine  the stochastic time evolution of the traffic flow in terms of the  incoherent classical hopping terms.
To do so, we formulate the incoherent (classical) dynamics of the open quantum system in terms of a Lindblad equation for the time evolution
of the density matrix $\rho ( t )$, describing  the interaction between different dots and between the boundary dots and a set of external reservoirs in terms of the jump, or Lindblad,
operators, $L_k$~\cite{Temme}:
\begin{eqnarray}
\dot{\rho}\left(t\right)&=&\sum_{k}\left(L_{k}\rho \left(t\right)  L_{k}^{\dagger}-\frac{1}{2}\left\{ L_{k}^{\dagger}L_{k},\rho \left(t\right) \right\} \right) \nonumber \\
&\equiv& \sum_{k} \mathcal{L}_k \left(\rho\left(t\right)\right)
\label{eq:lindbladeq} \, .
\end{eqnarray}

In our model we introduce two kinds of jump operators.  Operators of the first kind act locally on the boundary sites, creating and destroying vehicles;
instead, operators of the second kind  describe the incoherent stochastic transport of vehicles, thus  playing  a role similar to the Hamiltonian governing
the coherent transport in the Liouvillian equation. All these operators can be expressed in terms of the $\sigma_{i,j}^{(\ell)}$ operators.

The creation and annihilation Lindblad operators are defined as
\begin{eqnarray}
L_{{\rm in},j}^{(\ell)} & = & \sqrt{\Gamma_{j}^{(\ell)}} \sigma^{\left(\ell\right)}_{j,0}\nonumber \\
L_{{\rm out},j}^{(\ell)} & = & \sqrt{\gamma_{j}^{(\ell)}} \sigma^{\left(\ell\right)}_{0,j}
\end{eqnarray}
where $\Gamma_{j}^{(\ell)}$ and $\gamma_{j}^{(\ell)}$, with $\ell$ a boundary site, are the coupling strengths to inject or remove a vehicle.
The coupling constants give the number of vehicles that would like to enter or exit the system in the unit of time. $\Gamma_{j}^{(\ell)}$ is a
complicated function of the properties of the road (road surface, track width, and so on) and of the vehicle flux that would enter
the system. On the other hand, $\gamma_{j}^{(\ell)}$ only depends on the properties of the road, and for instance to get the flux of vehicles of
type $j$ exiting the system one should multiply $\gamma_{j}^{(\ell)}$ by the number of vehicles $\sigma^{\left( \ell \right)}_{j,j}$. We will
come back to the details on the physical meaning of these constants in the next Section, when we will discuss specific models.

Within our notation, we can easily introduce operators that move a given vehicle from the site $\ell$ to the neighbouring site $\ell'$,
given by
\begin{equation}
L^{(\ell,\ell')}_{{\rm hop},j,j'}=\sqrt{t_{j,j'}^{\left( \ell,\ell' \right)}}\sigma_{j',0}^{(\ell')}\sigma_{0,j}^{\left(\ell\right)} \, ,
\end{equation}
\noindent
with $t_{j,j'}^{\left( \ell,\ell' \right)}$ the coupling strength associated to the hopping process.
The Lindblad equation should contain only jump operators between neighboring sites in accordance with the correct driving directions
(a pedestrian can in principle move forward and backward between two dots while a vehicle can in general only move forward).

Having highlighted the general properties of the  Lindblad operators and of the  equations that they should satisfy, in the next Sections  we provide
their explicit expression  for  specific models, {\it i.e.} a three-dot model in Section \ref{cross} and a six-dot model in Section \ref{sixdot}.
Specifically, we will be focusing onto    the non-equilibrium steady state transport properties of these models, defined by the condition $\dot{\rho}=0$ and,
in particular, on the expectation values of the incoming and outgoing fluxes from and to the boundary reservoirs and the on site occupation levels.

\section{The three-dot minimal model}
\label{cross}

\begin{figure}
\center
\includegraphics*[width=0.6 \linewidth]{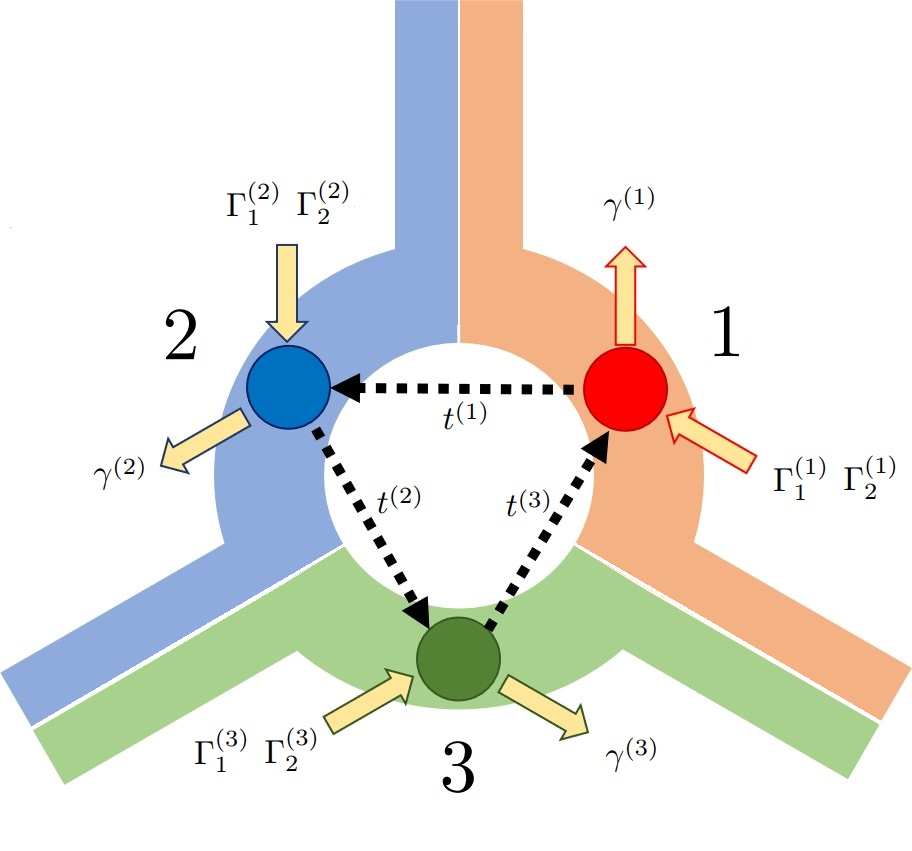}
\caption{Sketch of the minimal configuration for the open quantum crossroad. Each dot is associated to an entering road and to
the subsequent exit road, including the linking roundabout segment. We assume that, to each dot,
two monodirectional chains are coupled, injecting and removing vehicles with rates given by $\Gamma_j^{(\ell)}$ and
$\gamma^{(\ell)}$, respectively. The motion inside the roundabout is ruled by the hopping terms proportional to $t^{(\ell)}.$}
\label{minimal}
\end{figure}

In this Section we study a three-way crossroad, where the roads meet within a small roundabout.  In Fig.~\ref{minimal},
we depict our system, along with its minimal three-dot discretized version (in the next Section we will use a more realistic six-dot
discretization in order to properly implement priority rules). The crossroad has three entry and exit roads, labeled by $\underline{1}$,
$\underline{2}$, $\underline{3}$. Vehicles can enter and exit from each road. For the time being, we do not take into
account any priority rule:  as the only over-all constraint, we impose
 the exclusion principle, that is, the impossibility for two vehicles to occupy the same dot at the same time.
We consider a single type of vehicles, so that each site is described as an ($N$+1)-level dot with $N=3$, since three is the number of possible destinations.

In the minimal three-dot model each dot is labeled by indices $(1)$, $(2)$ and $(3)$, with the  dot $(\ell)$ representing  the incoming road
$\underline{(\ell-1)}$ and the outgoing road $\underline{(\ell)}$. On each dot we label the four possible states as: $\left|0\right\rangle$
for the empty site, $\left|1\right\rangle$ for a vehicle  that wants to exit at the first available exit, $\left|2\right\rangle$ for a vehicle
that wants to exit at the second available exit and $\left|3\right\rangle$ for a vehicle  that wants to exit through the same road it entered from.

 Assuming that there are no vehicles that go in and out from the same road, at each dot  we can totally remove from the Fock space the state $\left|3\right\rangle$.
 As a result, the Lindblad equation describing the roundabout includes now 12 jump operators:

\begin{itemize}

\item Three jump operators between consecutive dots, that convert the state $\left|2\right\rangle$ vehicle  on site $\ell $ into the state
$\left|1\right\rangle$ vehicle  on site $\ell +1$, as long as the destination site is empty. These  are given by (we removed inessential indices)
\begin{eqnarray}
L_{\rm hop}^{\left(1,2\right)} & = & \sqrt{t^{(1)}}\sigma_{1,0}^{\left(2\right)}\sigma_{0,2}^{\left(1\right)}\nonumber \\
L_{\rm hop}^{\left(2,3\right)} & = & \sqrt{t^{(2)}}\sigma_{1,0}^{\left(3\right)}\sigma_{0,2}^{\left(2\right)}\\
L_{\rm hop}^{\left(3,1\right)} & = & \sqrt{t^{(3)}}\sigma_{1,0}^{\left(1\right)}\sigma_{0,2}^{\left(3\right)}\nonumber
 ;
\end{eqnarray}

\item Six creating operators, two for each dot, injecting vehicles from the reservoirs and for each  available destinations
\begin{eqnarray}
L_{{\rm in},1}^{\left(1\right)}=\sqrt{\Gamma_{1}^{(1)}}\sigma_{1,0}^{\left(1\right)} & \ \ \ \  & L_{{\rm in},2}^{\left(1\right)}=\sqrt{\Gamma_{2}^{(1)}}\sigma_{2,0}^{\left(1\right)}\nonumber \\
L_{{\rm in},1}^{\left(2\right)}=\sqrt{\Gamma_{1}^{(2)}}\sigma_{1,0}^{\left(2\right)} & \ \ \ \  & L_{{\rm in},2}^{\left(2\right)}=\sqrt{\Gamma_{2}^{(2)}}\sigma_{2,0}^{\left(2\right)}\\
L_{{\rm in},1}^{\left(3\right)}=\sqrt{\Gamma_{1}^{(3)}}\sigma_{1,0}^{\left(3\right)} & \ \ \ \  & L_{{\rm in},2}^{\left(3\right)}=\sqrt{\Gamma_{2}^{(3)}}\sigma_{2,0}^{\left(3\right)}\nonumber
 ;
\end{eqnarray}
\item Three annihilation operator, removing vehicles in state $\left|1\right\rangle$ from the system
\begin{eqnarray}
L_{\rm out}^{\left(1\right)} & = & \sqrt{\gamma^{(1)}}\sigma_{0,1}^{\left(1\right)}\nonumber \\
L_{\rm out}^{\left(2\right)} & = & \sqrt{\gamma^{(2)}}\sigma_{0,1}^{\left(2\right)}\\
L_{\rm out}^{\left(3\right)} & = & \sqrt{\gamma^{(3)}}\sigma_{0,1}^{\left(3\right)}\nonumber
\:.
\end{eqnarray}
\end{itemize}
We refer to  Fig.~\ref{minimal} for a clearer definition of the symbols.

A subtle, though extremely important point is that, if we restrict our operator set only to the
12 operators listed above, as $t \to \infty$, the system would ultimately flow toward  an ``insulating'' phase. Indeed, none of
the operators above can change the state  $\left|2\right\rangle \otimes \left|2\right\rangle \otimes \left|2\right\rangle$,
that would stop the flux of vehicles. This effect is known to arise  in classical traffic flow when, at an intersection,
vehicles with the same priority meet at the same time. Clearly, in real life this situation is released by a collective
``smart'' behavior where drivers collaborate against a strict application of the driving rules (the so called
``courtesy crossing''). To emulate this behavior, and therefore to recover a non-zero and physically sound current across the intersection,
we need to introduce a 13th collective jump operator, that is, the

\begin{itemize}
\item Clockwise operator, that simultaneously move  three vehicles one step forward and is given by
\begin{eqnarray}
L_{\rm C} & = & \sqrt{\gamma_{\rm C}}\sigma_{1,2}^{\left(1\right)} \sigma_{1,2}^{\left(2\right)} \sigma_{1,2}^{\left(3\right)} \; , \label{clock}
\end{eqnarray}
\noindent
with $\gamma_{\rm C}$ the coupling strength for such collective process.
\end{itemize}
\noindent
Despite being a really simple system, the three-dot minimal model allows us  to derive useful traffic-flow information.
Defining the elementary moves of vehicles on the graph in terms of Lindblad operators and solving Eq.(\ref{eq:lindbladeq}) gives us full access to any kind of 
information on the system, at any time $t$ and in any point of the space.  For instance, we 
readily recover the single operator expectation values
\begin{equation}
\left\langle \hat{O} \left( t \right)  \right\rangle =\mathrm{Tr}\left(\hat{O}\rho\left( t \right)\right)\;, \label{single}
\end{equation}
as well as the  two (or more) operator expectation values~\cite{quantum_noise}
\begin{equation}
\left\langle \hat{O}_1 \left(t_1 \right) \hat{O}_2  \left( t_2 \right) \right\rangle=\mathrm{Tr}\left(\hat{O_1}e^{\left(t_1-t_2\right)\sum_k \mathcal{L}_k}\hat{O_2}\rho\left( t_2 \right)\right)\;, \label{two-time}
\end{equation}
where the operators  $\hat{O}_1$ and $\hat{O}_2$ can in principle act on different sites at times $t_1$ and $t_2$. 
Eqs.(\ref{single},\ref{two-time}) are the fundamental quantities for traffic forecasting and control~\cite{forecast_1,forecast_2}. Nevertheless, it is in general not easy to compute them 
analytically~\cite{correlation_1}, as the expectation values of any combination of $N$ operators typically depend on the expectation values of combinations of $N' \geq N$ operators. 
Because of that, the master equation approach usually relies on the mean field approximation that reduces the hierarchical set of equations to a closed set of $\mathcal{N}<\mathcal{D}$ nonlinear equations. 
For instance, the evolution equations for the expectation values of an operator acting on site $j$ depend only on single site operator expectation values~\cite{sudarshan}, i.e.
\begin{equation}
\left\langle \dot{\hat{O_{j}}}\left(t\right) \right\rangle   =f_{j}\left(\left\{ \left\langle \hat{O_{k}}\right\rangle \right\} _{k=1,\mathcal{N}}\right)\:\:,
\end{equation}
with $j=1,\ldots,\mathcal{N}$.
However, with the exception of some special cases~\cite{mpemba,selective_1,selective_2} where such an approximation becomes exact, the mean field approximation is not able to distinguish and compare different scenarios~\cite{is_2}.
For this reason in the following we prefer to solve the exact problem computing the full density matrix of the system. It is important to point out that in the density matrix approaches one is able to find the expectation values of the
 observables, for stochastic systems, as computed over the infinite number of possible evolutions starting from a given initial configurations (including the case in which the initial configuration corresponds to an ensemble and not to a pure state). 
 
 An alternative approach to the problem is provided by   numerical techniques, such as 
  the probabilistic cellular automaton models~\cite{ca_traffic_1}. With those methods, one typically  only computes one possible evolution at each run so that a proper average over multiple runs has to be done afterwards. Within our density matrix 
  approach, the average is intrinsically taken and, in fact, this is the reason for 
  which cellular automaton models look computationally less demanding. In fact, this is true, in general, for simple networks and one-operator expectation values, like Eq.(\ref{single}), where expectation values computed over few random
   configurations quickly approach the correct ensemble result. However, it is reasonable to expect that, increasing the complexity of the network and for two- (or more-) operator expectation values, like Eq.(\ref{two-time}), the number of 
   required configurations in the cellular automaton approach increases drastically. Furthermore, while cellular automaton approaches are particularly efficient for local interactions and instantaneous evolution rules (the evolution at time $t+\Delta t$ 
   depends on  the state of the system at time $t$), the density matrix approach complexity does not increase if we consider multisite evolution rules, like the one in Eq.(\ref{clock}), or if we replace the coupling strength with functions that 
    depend on the history of the system. For these reasons we conclude that a density matrix approach is more convenient when studying multiline intersections where nonlocal correlations between different directions play a crucial role~\cite{uta_1}.

The main relations  we are interested in are the formula for the occupation on site $\ell$, given by
\begin{equation}
\left\langle n_{\ell}\left( t \right)\right\rangle =1- \left\langle \sigma_{0,0}^{(\ell)} \left( t \right) \right\rangle \equiv 1- \mathrm{Tr}\left(\sigma_{0,0}^{(\ell)}\rho\left( t \right)\right)\;,
\end{equation}
and the balance equation for the occupation on site $\ell$, given by
\begin{equation}
\left\langle \dot{n_{\ell}} \left( t \right)\right\rangle =\mathrm{Tr}\left(n_{\ell} \dot{\rho}\left( t \right)\right)=\sum_k \mathrm{Tr}\left(n_{\ell} \mathcal{L}_k\left(\rho\left(t\right)\right) \right)  \;. \label{eq:UTA}
\end{equation}
We observe that each Lindblad operator gives rise to a current term of the form
\begin{equation}
I_k=\mathrm{Tr}\left(n_{\ell} \mathcal{L}_k\left(\rho\left(t\right)\right) \right) \;, \label{curr}
\end{equation}
that can be positive, if it describes vehicles that are entering into the site $\ell$, or negative, if it describe vehicles that are leaving the site $\ell$.
Explicitly separating such contributions, Eq.(\ref{eq:UTA}) can be written in the form
\begin{equation}
\left\langle \dot{n_{\ell}}  \left( t \right) \right\rangle=\sum_{k'} I^>_{k'} - \sum_{k"} I^<_{k"} \; ,
\label{balance}
\end{equation}
 where $>(<)$ describes entering(leaving) vehicles. Eq.(\ref{balance}) then  is formally equivalent to Eq.(22.5) of Ref.~\cite{uta_1} for the urban traffic analysis (UTA) model for traffic prediction
 in city networks. However, while the UTA model~\cite{uta_1,uta_2} is a macroscopic model where the currents are given in terms of effective flow rates, in our microscopic model the currents can be analytically
  computed at any time, $t$, through Eq.(\ref{curr}).
Of particular interest, are the incoming and outgoing currents that describe the vehicles flow between the intersection and the reservoirs. We define the total incoming flux through dot $\ell$ from the reservoir, given by
\begin{eqnarray}
\left\langle I_{{\rm in},\ell} \left( t \right)  \right\rangle  &=&\left\langle \left(\Gamma_{1}^{(\ell)}+\Gamma_{2}^{(\ell)}\right)\sigma_{0,0}^{(\ell)} \left( t \right) \right\rangle \nonumber \\
&\equiv& \mathrm{Tr}\left(\left(\Gamma_{1}^{(\ell)}+\Gamma_{2}^{(\ell)}\right)\sigma_{0,0}^{(\ell)}\rho\left( t \right)\right)\;,
\end{eqnarray}
and the outgoing flux through dot $\ell$ of vehicles of type $1$ to the reservoirs, given by
\begin{equation}
\left\langle I_{{\rm out},\ell} \left( t \right) \right\rangle  =\left\langle \gamma^{(\ell)}\sigma_{1,1}^{(\ell)} \left( t \right) \right\rangle  \equiv
\mathrm{Tr}\left(\gamma^{(\ell)}\sigma_{1,1}^{(\ell)}\rho \left( t \right)\right) \;.
\end{equation}
Other useful quantities we are interested in are the two-point and three-point density correlation, that is the joint probability for two or three sites to be occupied at the same time, $\left\langle n_{\ell}  \left( t \right) n_{k}  \left( t \right)\right\rangle$ and $\left\langle n_{\ell} \left( t \right)n_{k} \left( t \right)n_{p} \left( t \right)\right\rangle$.

\begin{figure}
\center
\includegraphics*[width=1. \linewidth]{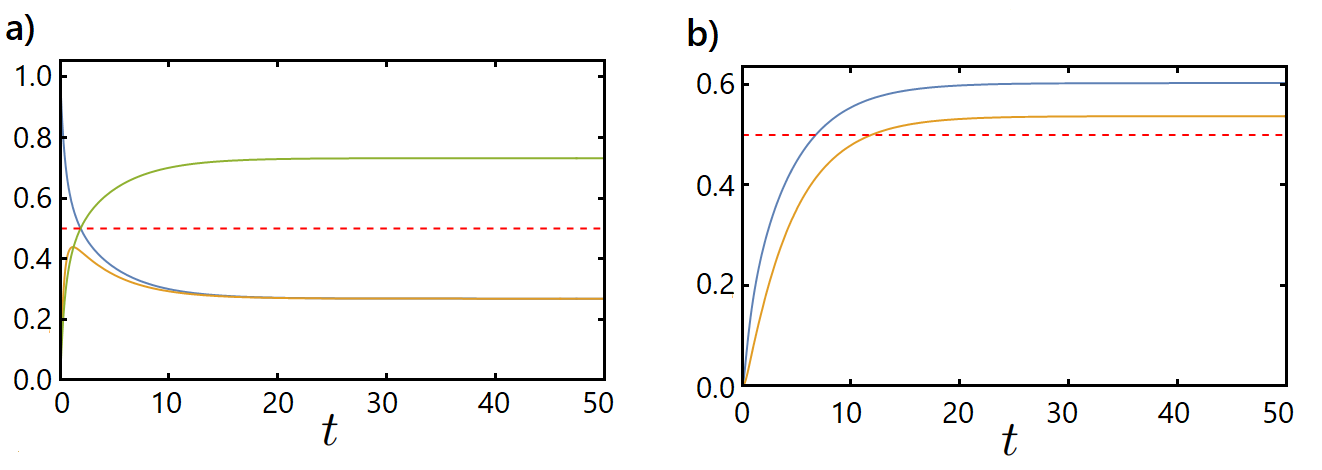}
\caption{Panel a) incoming flow at line 1 (blue curve), outgoing flow at line 1 (yellow curve), occupancy on dot 1 (green curve) and Panel b) $\langle n_1 n_2 \rangle$ (blue curve) and $\langle n_1 n_2 n_3 \rangle$ (yellow curve) for $\Gamma_j^{\ell}=0.5$ (dashed curve), $\alpha=0.5$, $\gamma_{\rm C}=0.1$, $t^{(\ell)}=1$.}
\label{time_gamma_1}
\end{figure}

\begin{figure}
\center
\includegraphics*[width=1. \linewidth]{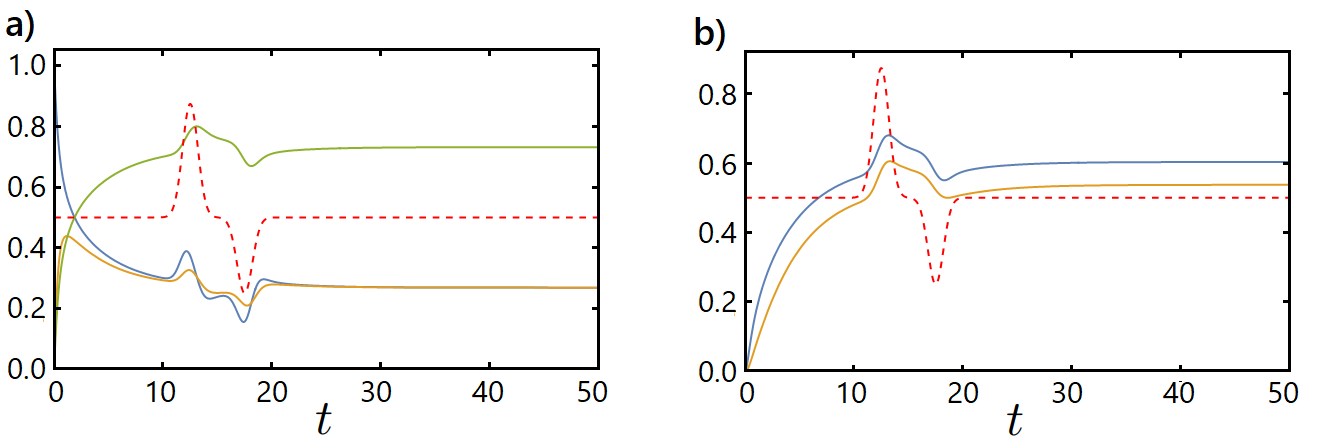}
\caption{Same as Fig.~\ref{time_gamma_1} but with different $\Gamma_j^{\ell}$ profile (dashed curve).}
\label{time_gamma_2}
\end{figure}

\begin{figure}
\center
\includegraphics*[width=0.5 \linewidth]{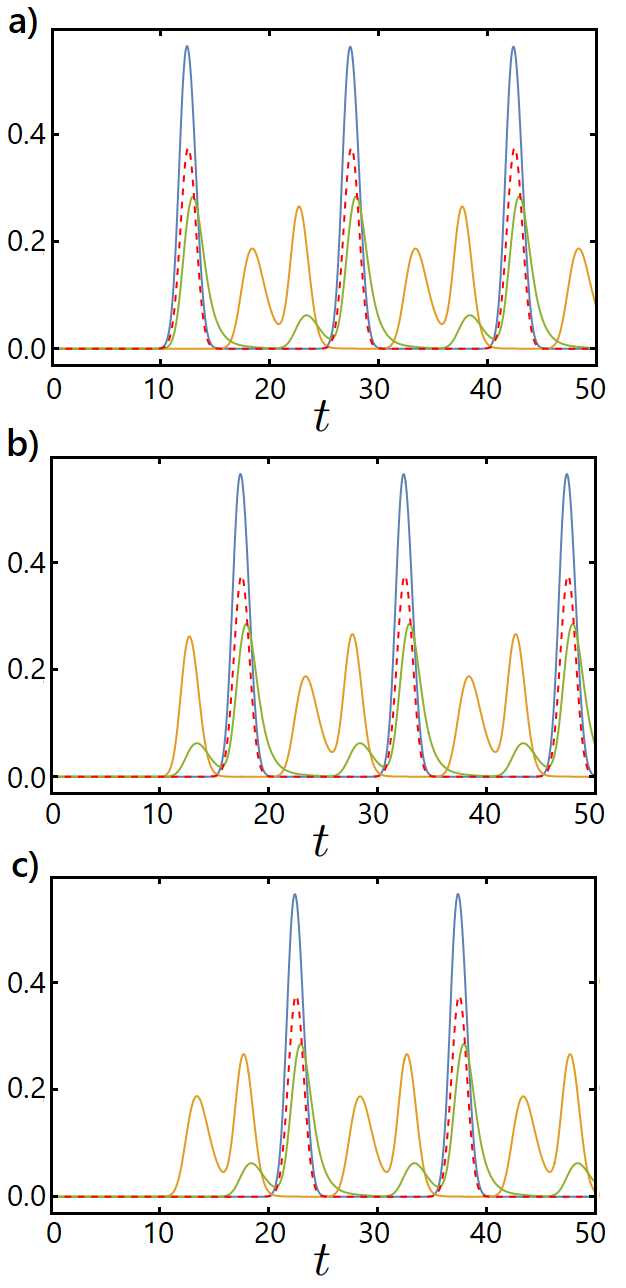}
\caption{Same as panel a) of Fig.~\ref{time_gamma_1} but with different $\Gamma_j^{\ell}$ profile (dashed curves) periodic and shifted for each line. We show incoming flow (blue curve), outgoing flow (yellow curve), occupancy (green curve) and incoming coupling strength (dashed line) for site 1 (panel a), site 2 (panel b) and site 3 (panel c).}
\label{time_gamma_3}
\end{figure}

In Figs.~\ref{time_gamma_1}, \ref{time_gamma_2} and \ref{time_gamma_3} we show the time evolution of the currents, the occupations and the density-density correlations for our system for different time depending values of the coupling strengths \footnote{We solved the Lindblad equation in Eq.\ref{eq:lindbladeq} for the full density matrix of the system making use of standard Mathematica functions as NDSolve for time integration and NSolve for the non-equilibrium steady state solution ($\dot\rho=0$). However, we plan to implement the code in Fortran or Python in a further development of our work, in order to reduce the computational cost.}. In particular, we 
assume $\gamma^{(\ell)}=3$, $\gamma_C=0.1$, $t^{(\ell)}=1$ and to start with an empty intersection at time $t=0$ with the values of the $\Gamma_k^{(\ell)}$ reported on the plots (dashed lines). In Fig.~\ref{time_gamma_1} we assume constant incoming coupling strength 
$\Gamma_j^{(\ell)}=0.5$, in this case all the observables quickly reach a stationary state at $t\approx20$. In Fig.~\ref{time_gamma_2} we assume that above a constant value of $\Gamma_j^{\ell}=0.5$ we have a sudden increase at $t\approx 12.5$ and a sudden decrease 
at $t \approx 17.5$ of the incoming coupling strengths. From the plot we can observe how the system reacts during and after the variation of the coupling strength before reaching the stationary state. Finally in Fig.~\ref{time_gamma_3} we consider the case of periodic 
coupling strengths dephased between each line, i.e. a traffic light. In this case the system never reaches a stationary state but exhibits a periodic behavior. Let us note that the shift in time between outgoing and incoming fluxes at each line is a consequence of the intrinsic exclusion principle of the Fock space that emerges from the conditions on the $\sigma_{i,j}^{(\ell)}$ operators.

From now on, we assume the coupling strengths to be time-independent real numbers so that we will focus
on the non-equilibrium steady state values defined by setting  $\dot{\rho}=0$.

In order to reduce the number of parameters, we now consider a $\mathbb{Z}_{3}$-symmetric roundabout, by setting $\gamma^{(\ell)}=\gamma$,
$\Gamma_{1}^{(\ell)}=\alpha\Gamma$, and $\Gamma_{2}^{(\ell)}=(1-\alpha)\Gamma$,
$\forall \ell=1,2,3$, with $\alpha$ that regulates the mismatch between vehicles that want to exit
at the first or second available exit.

\begin{figure}
\center
\includegraphics*[width=0.8 \linewidth]{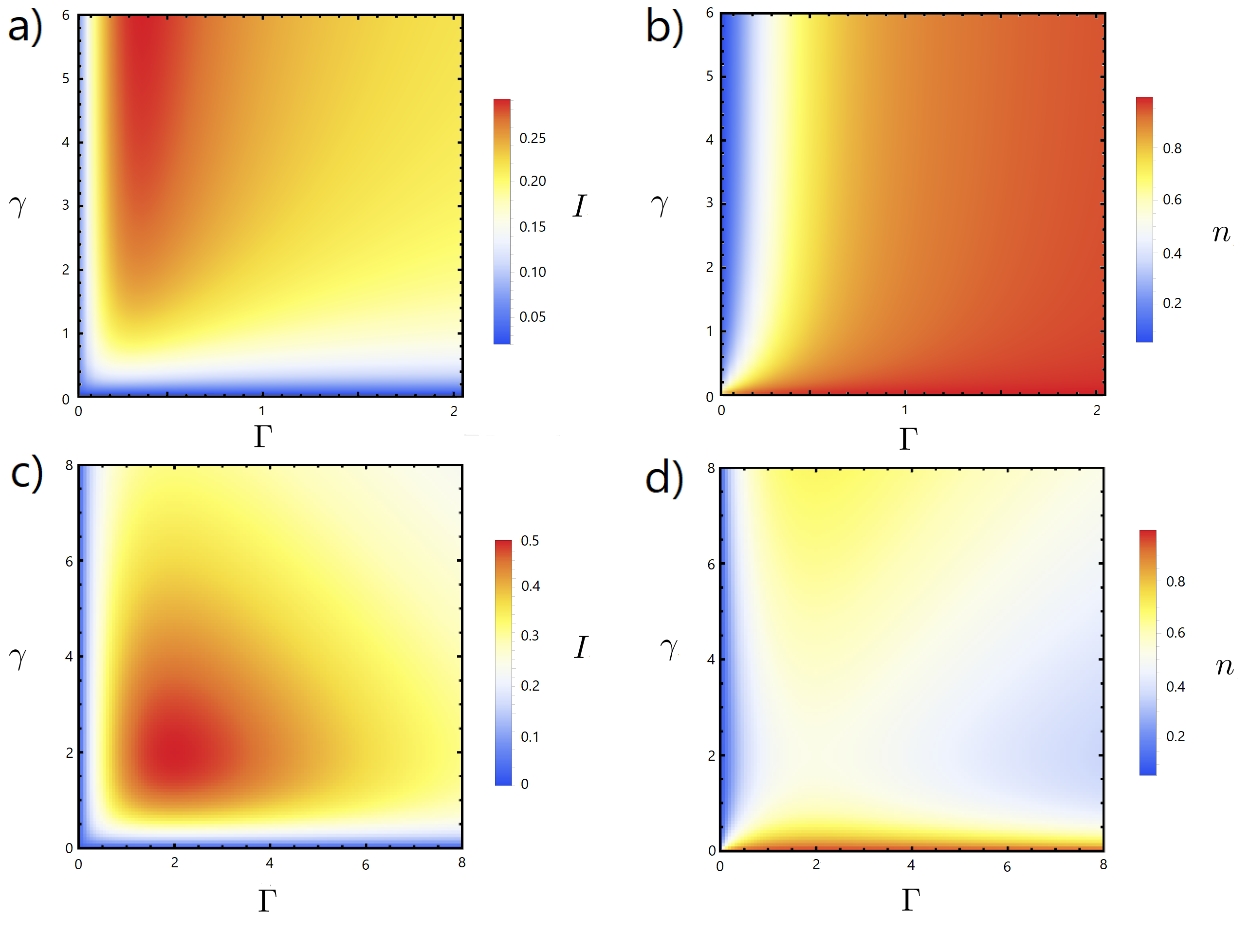}
\caption{Total current across the junction (panel a) and mean occupation number (panel b) as a function of the incoming, $\Gamma$, and outgoing, $\gamma$, couplings for $\alpha=0.5$, $\gamma_{\rm C}=0.1$, $t^{(\ell)}=1$. Panel c) and d) show the same quantities as in panel a) and b) but for the fully quantum fermionic chain discussed in Ref.~\cite{NRG}.}
\label{current-density}
\end{figure}
\noindent
Let $I_{{\rm in} , \ell}$ and $I_{{\rm out} , \ell}$ respectively denote the current ({\it i.e.} the flux) of vehicles entering or exiting the
roundabout at route $\ell$.
In Fig.~\ref{current-density} we show   the steady state  current across the junction, $I=\left(I_{{\rm in},1}+I_{{\rm in},2}+I_{{\rm in},3} \right)/3
= \left( I_{{\rm out},1}+I_{{\rm out},2}+I_{{\rm out},3} \right)/3$, and the total mean density, $n=\left( n_1+n_2+n_3 \right)/3$.
 In drawing the plots, we have set a finite cooperative coupling $\gamma_{\rm C}=0.1$ and we have assumed  an equal number of vehicles for both exit directions,
that is, $\alpha=0.5$. We measure everything in units of  $t^{(\ell)}\equiv 1$.
Two interesting behaviors emerge in both the  current and the density. The current exhibits an optimal working point in the $\Gamma$ direction,
similar to the one that is observed in one-dimensional  quantum fermionic, or spin, chains (see for instance, Fig.~3a of Ref.~\cite{NRG}, here reproduced in panel c) of Fig.~\ref{current-density});
however, the quantum optimal working point in the $\gamma$ direction is spoiled in the classical incoherent case we are treating here.
Furthermore, while  in a quantum chain the  optimal working point is in pair with a similar non-monotonic behavior of the mean density
(see Fig.~4 of Ref.~\cite{NRG}, here reproduced in panel d) of Fig~\ref{current-density}), in the classical regime this is not the case and the density increases monotonically as
a function of $\Gamma$, for any value of $\gamma$. The reason is that, in the coherent quantum case, the optimal working point
emerges as a consequence of the bulk-reservoir hybridization, while in the incoherent classical dynamics it is a consequence of
the exclusion statistic at the junction that generates a competition between vehicles that want to occupy the same dot from different
sites. We also observe, as expected, that both current and density are independent of  $\gamma$
for high enough values. This is due the absence of backscattering terms (that are instead present in the quantum case of Ref.~\cite{NRG} due the hermitianity condition on the system Hamiltonian): vehicles are allowed to move in only one direction, so,
as soon as $\gamma$ is strong enough in order to sink all the incoming vehicles without delay, it has no other effects on the junction.

\begin{figure}
\center
\includegraphics*[width=0.5 \linewidth]{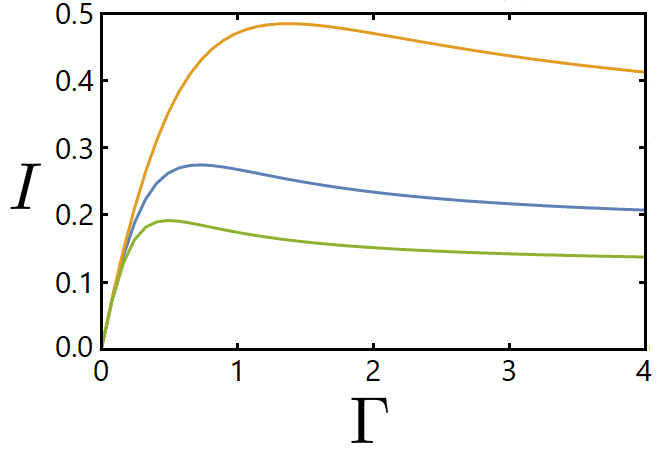}
\caption{Total current across the junction and mean occupation number as a function of $\Gamma$ for $\gamma=3$, $\gamma_{\rm C}=0.1$, $t^{(\ell)}=1$ and
$\alpha=0.25$ (green curve), $\alpha=0.5$ (blue curve) and $\alpha=0.75$ (orange curve).}
\label{current3}
\end{figure}

\begin{figure}
\center
\includegraphics*[width=0.5 \linewidth]{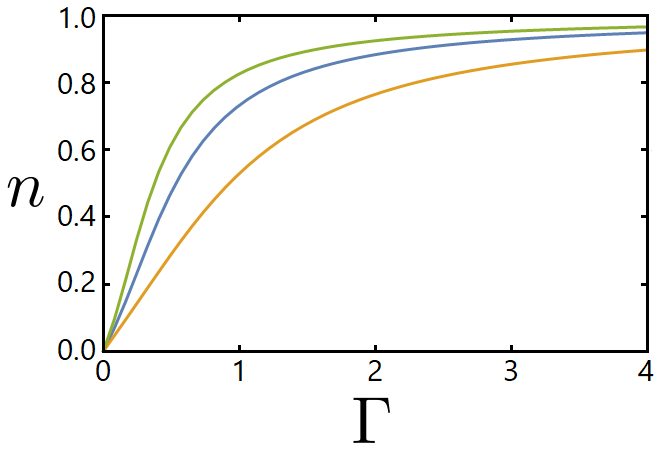}
\caption{Total density across the junction and mean occupation number as a function of $\Gamma$ for $\gamma=3$, $\gamma_{\rm C}=0.1$, $t^{(\ell)}=1$ and
$\alpha=0.25$ (green curve), $\alpha=0.5$ (blue curve) and $\alpha=0.75$ (orange curve).}
\label{density3}
\end{figure}

\begin{figure}
\center
\includegraphics*[width=0.5 \linewidth]{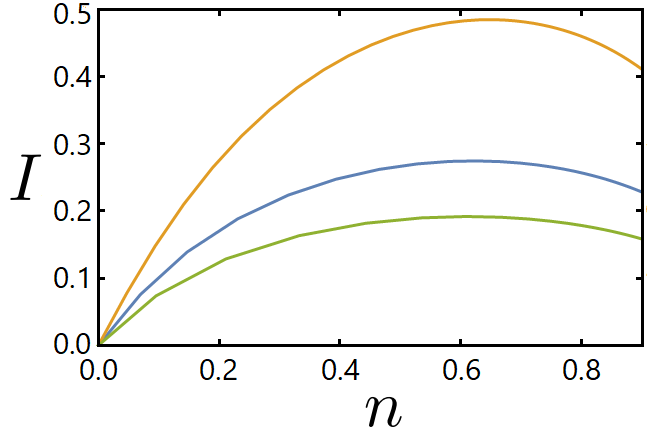}
\caption{Fundamental traffic-flow diagram (traffic flux $I$ as a function of the mean occupation $n$)
for $\gamma=3$, $\gamma_{\rm C}=0.1$, $t^{(\ell)}=1$ and $\alpha=0.25$ (green curve), $\alpha=0.5$ (blue curve) and $\alpha=0.75$ (orange curve).}
\label{foundamental}
\end{figure}

\noindent
The nonmonotonic behavior in the current and the monotonic one in the density are  a footprint  of the classical
traffic-jam transition. In Fig.~\ref{current3} and in
Fig.~\ref{density3}, we respectively show the dependence of the current and of the density on $\Gamma$. Combining
the two figures, we draw  Fig.~\ref{foundamental}, where  we show the current as a function of the density for $\gamma=3$ and $\alpha=0.25, 0.5, 0.75$
recovering the so called fundamental traffic-flow diagram, that is the existence of a critical density value beyond which a traffic-jam phase
occurs, with a consequent decrease in the current.  In Fig.~\ref{foundamental} we see a net increase of the current as a function of $\alpha$.
In fact, we expect a trend as such, as a consequence of the fact that, the larger is $\alpha$ (the closer to 1), the less the vehicles from
different branches compete with each other for the same site occupation, till they do not compete at all for $\alpha = 1$.

While the results we show in Figs.~\ref{current3}, ~\ref{density3} and ~\ref{foundamental} have been derived for the $\mathcal{Z}_3$ symmetric case,
when generalizing them to  the asymmetric case, we readily infer that low/high values of the $\Gamma_j^{(\ell)}$ are associated to a low/high-density
regime, with the current that exhibits a maximum for a finite intermediate value of the coupling strengths.

It is worth to note that, even if we do not explicitly introduce in our model the delay time, which is a fundamental parameter of
the car-following microscopic models based on continuum space-time description~\cite{Bando}, we are  still able to recover a fundamental  traffic-flow diagram describing the jamming transition from the free flow to the congested regime. This is not surprising as in stochastic space-discrete models, like the one we implemented, the role of the delay time is played by a combined effect of the exclusion principle (that does not allow vehicles to jump in occupied sites)
and of the probabilistic nature of the hopping events (mimicking the vehicle velocity changes)~\cite{Nagel,Wagner,Siqueira}. While these effects are already present in the case of a one dimensional lane~\cite{Derrida,rossini,NRG}, in our model the traffic-jam transition emerges due the interaction between vehicles from different lanes and is then a direct consequence of the topology of the intersection.

So far,  we have not implemented any priority rule at the crossway, that is, we have assumed a sort of ``wild'' driving style, where any occupancy
competition is stochastically resolved by the ``fastest'' vehicle. However, in real-life experience, we know that the
wild driving style can be effective in the low-density case, but becomes ineffective and even dangerous with increasing number of vehicles.
For this reason, different kinds of priority rules have to be considered and implemented: right-way priority,
roundabout, presence of a traffic light, {\it et cetera}. Each of these rules has its own regime in which it works better than the others.
We can recover and predict such regimes also in our quantum dot model, properly implementing priority
rules into the jump operators. In order to deal with different priority rules on the same footing and compare them with each other,
in the next Section we   move from a minimal three-dot model to a more appropriate six-dot model, which we describe below.

\section{The six-dot model}
\label{sixdot}

\begin{figure}
\center
\includegraphics*[width=0.5 \linewidth]{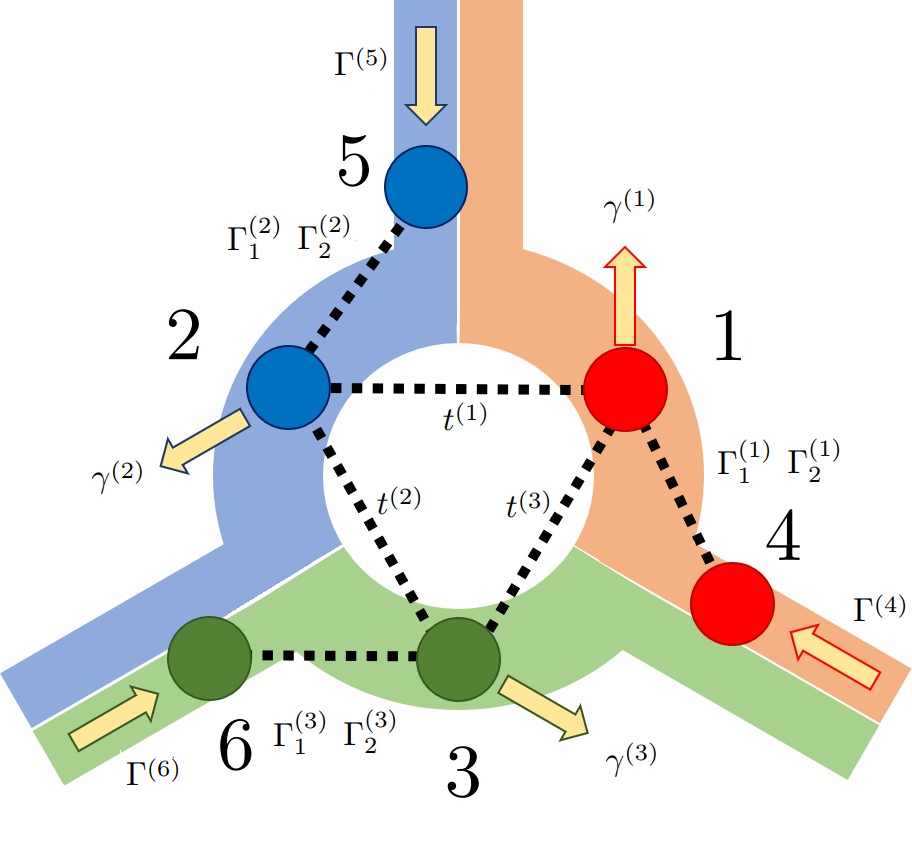}
\caption{Sketch of the six-dot configuration for the open quantum crossroad. Compared to the minimal model in Fig.~\ref{minimal},
we added three external sites. The baths that inject vehicles are now coupled to the external dots, with a flux proportional to
$\Gamma^{(\ell)}$, $\ell=4,5,6$, while the outgoing baths are coupled to the internal sites, with a flux proportional to
$\gamma^{(\ell)}$, $\ell=1,2,3$. The decision about the exit is made before entering into one of the internal dots of the
roundabout: this decision is governed by the coupling constants $\Gamma_j^{(\ell)}$, $\ell=1,2,3$, $j=1,2$.}
\label{6dots}
\end{figure}

The six-dot model is built out of  the three-dot one with the addition of three two-level extra sites. The internal sites
$\ell=1,2,3$ regulate the interaction between vehicles, exactly like in the three-dot model, while the external sites
$\ell=4,5,6$ regulate the injection of vehicles into the system. In absence of priority rules, a case we have dubbed ``wild case'',
the Lindbladian dynamics is described by the following jump operators, pictorially shown in Fig.~\ref{6dots}:

\begin{itemize}

\item The three jump operators between consecutive dots that convert a $\left|2\right\rangle$ vehicle on site $\ell$ into a
$\left|1\right\rangle$ vehicle on site $\ell+1$, with $\ell=1,2,3$,  do not change, as long as the arrival site is empty:

\begin{eqnarray}
L_{\rm hop}^{\left(1\right)} & = & \sqrt{t^{(1)}}\sigma_{1,0}^{\left(2\right)}\sigma_{0,2}^{\left(1\right)}\nonumber \\
L_{\rm hop}^{\left(2\right)} & = & \sqrt{t^{(2)}}\sigma_{1,0}^{\left(3\right)}\sigma_{0,2}^{\left(2\right)} \label{inthop}\\
L_{\rm hop}^{\left(3\right)} & = & \sqrt{t^{(3)}}\sigma_{1,0}^{\left(1\right)}\sigma_{0,2}^{\left(3\right)}\nonumber ;
\end{eqnarray}

\item The six creation operators, injecting vehicles from the reservoirs, now act on the
external sites: therefore, only three of them are left out and they create a vehicle without giving directional preferences:

\begin{eqnarray}
L_{\rm in}^{\left(4\right)}=\sqrt{\Gamma^{(4)}}\sigma_{1,0}^{\left(4\right)} \nonumber \\
L_{\rm in}^{\left(5\right)}=\sqrt{\Gamma^{(5)}}\sigma_{1,0}^{\left(5\right)} \\
L_{\rm in}^{\left(6\right)}=\sqrt{\Gamma^{(6)}}\sigma_{1,0}^{\left(6\right)} \nonumber ;
\end{eqnarray}

\item In addition,  we introduce six new hopping operators, that allow a vehicle to jump from an external to the corresponding internal
dot. During this process the decision about the exit is made: a vehicle on the external dot $\ell +3$ can decide to become a type-$\left| 1 \right\rangle$ , or a type-$\left| 2 \right\rangle$  vehicle, on site $\ell$:
\begin{eqnarray}
L_{{\rm hop},1}^{\left(4\right)}=\sqrt{\Gamma_1^{(1)}}\sigma_{1,0}^{\left(1\right)}\sigma_{0,1}^{\left(4\right)} & \
& L_{{\rm hop},2}^{\left(4\right)}=\sqrt{\Gamma_2^{(1)}}\sigma_{2,0}^{\left(1\right)}\sigma_{0,1}^{\left(4\right)} \nonumber \\
L_{{\rm hop},1}^{\left(5\right)}=\sqrt{\Gamma_1^{(2)}}\sigma_{1,0}^{\left(2\right)}\sigma_{0,1}^{\left(5\right)} & \
& L_{{\rm hop},2}^{\left(5\right)}=\sqrt{\Gamma_2^{(2)}}\sigma_{2,0}^{\left(2\right)}\sigma_{0,1}^{\left(5\right)} \label  {wildhop} \\
L_{{\rm hop},1}^{\left(6\right)}=\sqrt{\Gamma_1^{(3)}}\sigma_{1,0}^{\left(3\right)}\sigma_{0,1}^{\left(6\right)} & \
& L_{{\rm hop},2}^{\left(6\right)}=\sqrt{\Gamma_2^{(3)}}\sigma_{2,0}^{\left(3\right)}\sigma_{0,1}^{\left(6\right)} \nonumber ;
\end{eqnarray}

\item The three annihilation operators removing vehicles in the state $\left|1\right\rangle$ from an internal site of the junction do not change:

\begin{eqnarray}
L_{\rm out}^{\left(1\right)} & = & \sqrt{\gamma^{(1)}}\sigma_{0,1}^{\left(1\right)}\nonumber \\
L_{\rm out}^{\left(2\right)} & = & \sqrt{\gamma^{(2)}}\sigma_{0,1}^{\left(2\right)}\\
L_{\rm out}^{\left(3\right)} & = & \sqrt{\gamma^{(3)}}\sigma_{0,1}^{\left(3\right)}\nonumber .
\end{eqnarray}

\item Clockwise operator, that simultaneously move  three vehicles one step forward, continues to act on the internal sites and is given by
\begin{eqnarray}
L_{\rm C} & = & \sqrt{\gamma_{\rm C}}\sigma_{1,2}^{\left(1\right)} \sigma_{1,2}^{\left(2\right)} \sigma_{1,2}^{\left(3\right)} \; .
\end{eqnarray}
\end{itemize}

\begin{figure}
\center
\includegraphics*[width=0.6 \linewidth]{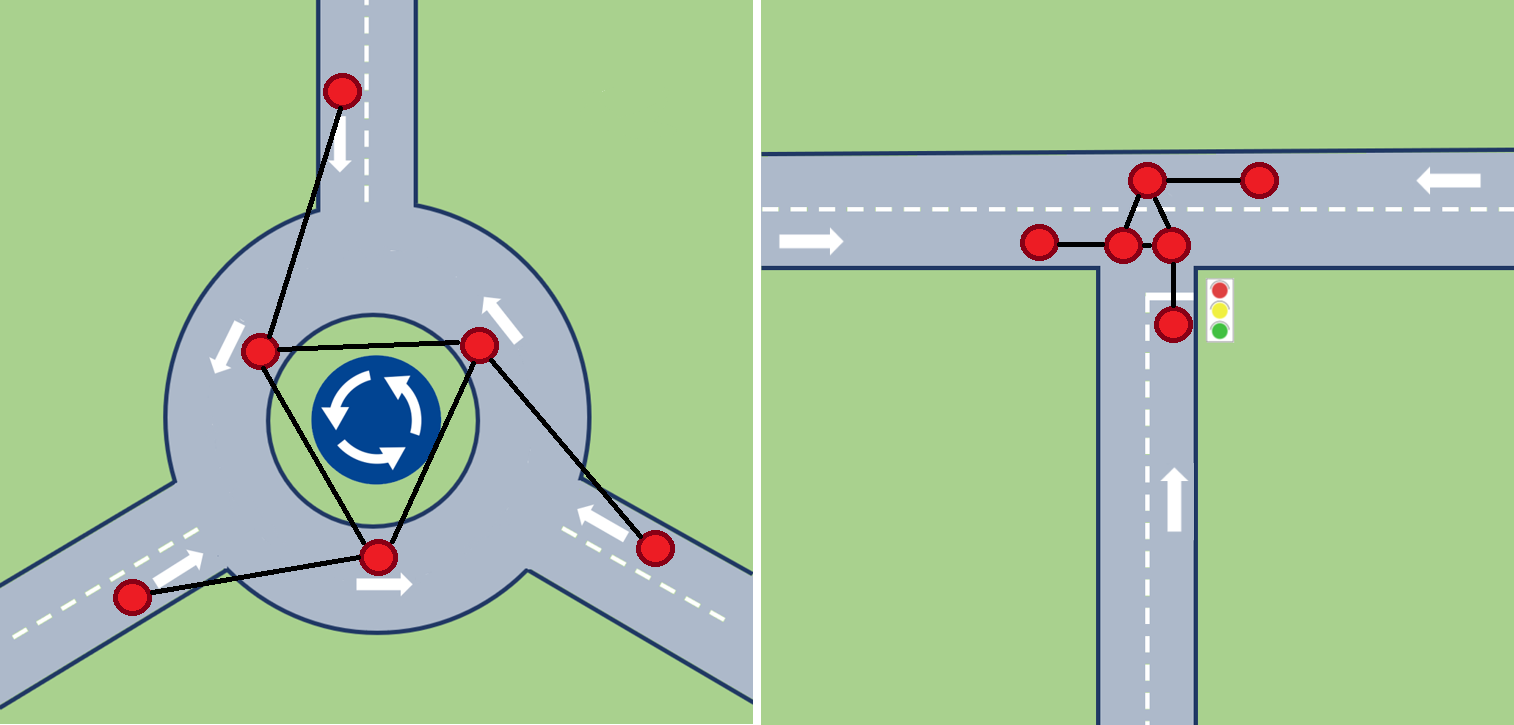}
\caption{Possible three-road intersection discussed within the six-dot model. Both intersections share the same topology,
but they differ for the imposed priority rules. In the roundabout, vehicles inside the ring have the priority over vehicles
from the external arms; in the symmetric intersection, right-hand priority rule applies; in the priority intersection,
vehicles in the secondary arm must give way to the vehicles from both major roads.}
\label{cases}
\end{figure}

\noindent
In the following, we implement various priority rules, on the background of the sets of operators listed above, by modifying the hopping
operators in Eq.(\ref{inthop}) or in Eq.(\ref{wildhop}). In addition, pertinently modifying the various hopping operators, we will be
able to
describe and compare on the same footing different road configurations, like the ones shown in Fig.~\ref{cases}.

\subsection{The roundabout}
\label{round}

If the intersection is realized by means of a small roundabout, vehicles inside the roundabout must have the priority over vehicles
coming from outside. In terms of jump operators, we implement this   by modifying  the operators in
Eq.(\ref{wildhop}) as follows

\begin{eqnarray}
L_{{\rm hop},j}^{\left(4\right)} \rightarrow \sqrt{\Gamma_j^{(1)}}\sigma_{j,0}^{\left(1\right)}\sigma_{0,1}^{\left(4\right)}
\left( \sigma_{0,0}^{\left(3\right)}+\beta \sigma_{1,1}^{\left(3\right)} \right) \nonumber \\
L_{{\rm hop},j}^{\left(5\right)} \rightarrow \sqrt{\Gamma_j^{(2)}}\sigma_{j,0}^{\left(2\right)}\sigma_{0,1}^{\left(5\right)}
\left( \sigma_{0,0}^{\left(1\right)}+\beta \sigma_{1,1}^{\left(1\right)} \right)    \label {correcthop}  \\
L_{{\rm hop},j}^{\left(6\right)} \rightarrow \sqrt{\Gamma_j^{(3)}}\sigma_{j,0}^{\left(3\right)}\sigma_{0,1}^{\left(6\right)}
\left( \sigma_{0,0}^{\left(2\right)}+\beta \sigma_{1,1}^{\left(2\right)} \right)\nonumber \, ,
\end{eqnarray}
\noindent
with $\beta$ measuring the promptness of the drivers to use the turn signals. If $\beta=1$, all the drivers
correctly use the lights to announce their direction. In this way a vehicle can jump into the $\ell$-th roundabout dot from the external
$\ell+3$  dot ($\ell =1,2,3$) if and only if the $(\ell-1)$ [mod 3] site is not occupied by a
type-$\left| 2 \right\rangle$ vehicle. On the other hand, when $\beta=0$, an incoming vehicle on site $\ell+3$
does not know whether a vehicle in site $\ell-1$ [mod 3] wants to jump, so it has to give the priority if
the site is occupied by either a type-$\left| 1 \right\rangle$-type, or  a type-$\left| 2 \right\rangle$ vehicle.

We consider a three-way intersection with two major fluxes ($\ell=4,5$) and a minor flux ($\ell=6$), setting
$\Gamma^{(4)}=\Gamma^{(5)}=\Gamma_>$ and $\Gamma^{(6)}=\Gamma_<$. The other coupling constants are chosen
consistently with the $\mathcal{Z}_3$
symmetry, that is we set $\gamma^{(\ell)}=\gamma$, $t^{(\ell)}=t$, with unbalanced exit-intention flux,
$\Gamma_{1}^{(\ell)}=\alpha^{(\ell)} t$, $\Gamma_{2}^{(\ell)}=(1-\alpha^{(\ell)}) t$ with  $\ell=1,2,3$.
To build a phase diagram of the junction we consider two cases:  the `wild case', {\it i.e.} the case with hopping
operators as in Eq.(\ref {wildhop}), and the `perfect-roundabout case', in which the hopping terms are given by Eq.(\ref {correcthop}) with
$\beta =1$. In both cases we compute the non-equilibrium steady state total current across the junction as a
function of $\Gamma_>$ and $\Gamma_<$, with $\Gamma_< < \Gamma_>$.

\begin{figure}
\center
\includegraphics*[width=0.7 \linewidth]{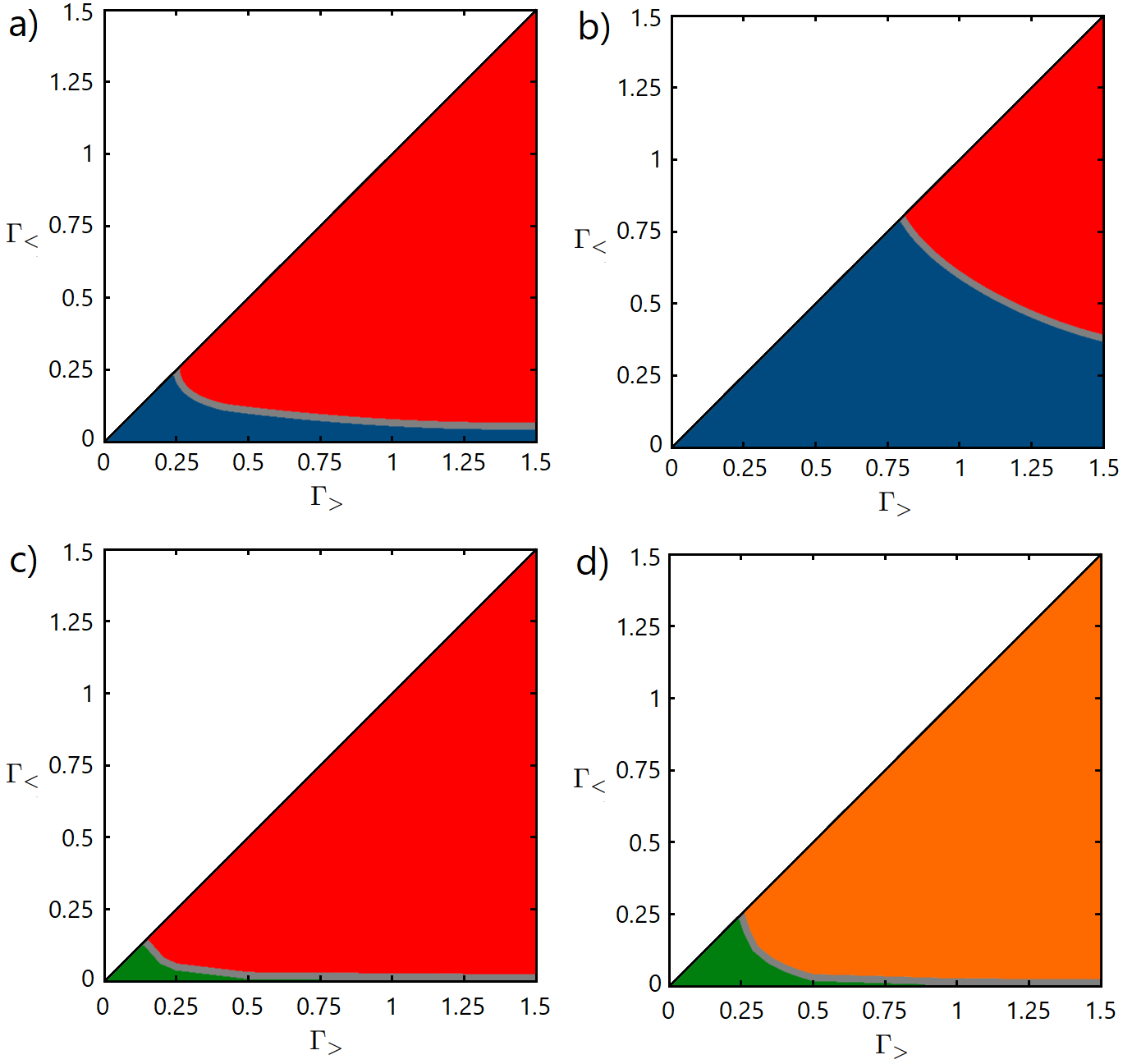}
\caption{Panel a): Phase diagram of a three-way roundabout with two major fluxes, $\Gamma^{(4)}=\Gamma^{(5)}=\Gamma_>$, and
a minor flux, $\Gamma^{(6)}=\Gamma_<$ and with constants $\gamma $ and $t$  $\mathcal {Z}_3$ symmetric.
At each point we plot which priority rule, between the wild one (blue region) and the perfect roundabout (red region),
exhibits the higher current across the system. We assume a symmetry between incoming vehicles that want to exit at the
first and second available exit by setting $\alpha ^{(\ell)} =0.5$, with $\ell=1,2,3$. The other parameters are set to
$t^{(\ell)}=1$, $\gamma_{\rm C}=0.1$, $\Gamma_j^{(\ell)}=0.5$, $\gamma^{(\ell)}=3$.
Panel b): Same as panel a) but with an asymmetry in the exit direction that favors the first available gate, $\alpha=0.75$.
Panel c): Same as panel b) but now we compare the right-hand priority rule (green region) and the perfect roundabout ($\beta =1$ one (red region).
Panel d): Same as panel b) but now we compare the right-hand priority rule (green region) and the no-turn signal roundabout ($\beta =0$ one (orange region).}
\label{merged}
\end{figure}

\noindent
In Fig.~\ref{merged}, panels a) and b), we show the regions in the plane
$\Gamma_<,\Gamma_>$ in which the wild or perfect roundabout priority rules produce the greatest current in the junction
 for $\alpha^{(\ell)}=0.5$ and $\alpha^{(\ell)}=0.75$, with $\ell=1,2,3$.
We observe that, for low values of $\Gamma_<$ and $\Gamma_>$, corresponding to a low-density regime, the wild case exhibits a greater
current than the perfect roundabout one; in contrast, for higher values of $\Gamma_<$ and $\Gamma_>$, that is in the high-density regimes, the perfect roundabout priority rule is more convenient.
We also note that increasing $\alpha^{(\ell)}$ enhances the region in which the wild case shows a current greater than in the perfect roundabout case.

\begin{figure*}
\center
\includegraphics*[width=0.9 \linewidth]{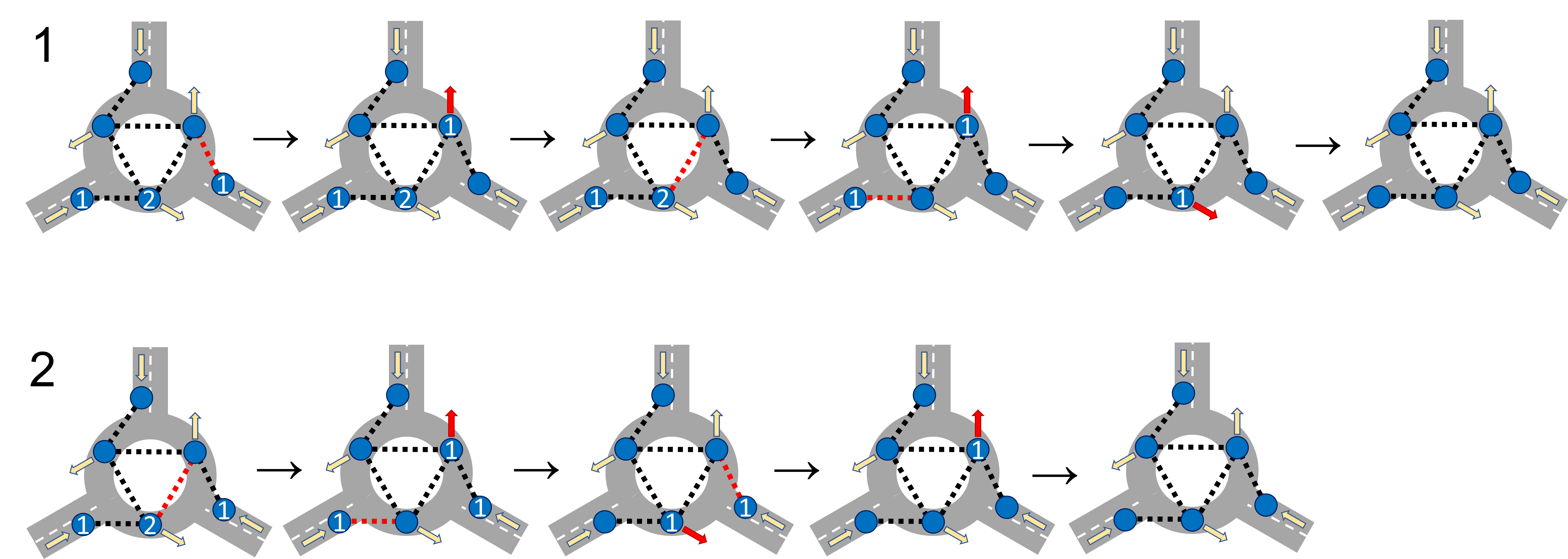}
\caption{Minimum number of turns, for a sample configuration, to allow all the vehicles to exit from the system for
a given priority rule. On the first line, we assume  right-hand priority, which requires 6 different  turns. On the second line,
we assume perfect
roundabout priority and, in this case,  five different turns are  sufficient. Number $2$ represent a vehicle that wants to exit at the
second possible gate, number $1$ a vehicle intentioned to exit at the first available gate. At each step we draw  in red color
the operator which is acting.}
\label{steps}
\end{figure*}

All our results can be explained in terms of classical behavior. Let us start from the high-density regime. First of all, let
us note that in the configuration $\left| 1\right\rangle\otimes \left| 1\right\rangle \otimes \left| 1\right\rangle$ vehicles do not
compete with each other and can freely enter into-,  and exit from-, the roundabout all together. This suggests that, getting rid of $\left| 2\right\rangle$-states
from the internal sites should increase the current. Furthermore, the probability to have two
consecutive dots both occupied increases with $\Gamma_>$ and $\Gamma_<$, as more vehicles are injected into the system
from the reservoirs. In Fig.~\ref{steps} we consider what happens to a specific configuration, involving more than one vehicle,
if the priority is given to the vehicle  inside the roundabout or to the vehicle from outside. Clearly,  in the roundabout configuration only
the first case happens. At variance,  in the wild case both possibilities are realized, that is, vehicles can take the priority regardless
of whether they are inside, or outside the junction. Now, as a general remark, we note that,  if the vehicles inside the junctions
have the priority, vehicles leave the junction one turn earlier (this characteristic
is quite general, as we have found the same behavior also for other configurations with ``many'' vehicles).
As such configurations are more and more likely at higher value of the density, we expect the roundabout configuration
to win against the wild one at high values of $\Gamma_>$ and $\Gamma_<$. Instead, in the complementary
low-density regime, that is at low values of the $\Gamma_>$ and $\Gamma_<$ couplings, a configuration like
the one in Fig.~\ref{steps} becomes unlikely, as only few vehicles will be present into the system at the same time.
On the other hand, we have to remember that the hopping probability is a stochastic event. Clearly, if the priority
rule allows only one of the vehicle to ``toss the coin" while the other ``must" wait, as in the roundabout case, the
risk is greater and, consequently the flux is smaller than in the case in which both vehicles are allowed to ``toss the coin".
For this reason we expect, as we  explicitly show within our model, that in the low-density regime the wild configuration is
more convenient in terms of current flow with respect to a perfect roundabout.

\subsection{Symmetric intersection: right-hand priority}
\label{right}

 In the previous Subsection we have compared the wild (no priority) setup with the roundabout priority rule. While the wild setup
is not realistic in real-life traffic configurations, due the high probability of accidents, the roundabout is not the only
commonly used  priority rule. Indeed, before its introduction, the right-hand (or left-hand) priority rule was the most commonly used one. It is then
natural to compare in our model the roundabout with, say, the right-hand priority rule. In order to simulate the right-hand priority
we only have to modify the operators in Eq.(\ref {inthop}) as follows:

\begin{eqnarray}
L_{\rm hop}^{\left(1\right)} & \rightarrow & \sqrt{t^{(1)}}\sigma_{1,0}^{\left(2\right)}\sigma_{0,2}^{\left(1\right)}\sigma_{0,0}^{\left(5\right)}  \nonumber \\
L_{\rm hop}^{\left(2\right)} & \rightarrow & \sqrt{t^{(2)}}\sigma_{1,0}^{\left(3\right)}\sigma_{0,2}^{\left(2\right)}\sigma_{0,0}^{\left(6\right)} \label{righthand}\\
L_{\rm hop}^{\left(3\right)} & \rightarrow & \sqrt{t^{(3)}}\sigma_{1,0}^{\left(1\right)}\sigma_{0,2}^{\left(3\right)}\sigma_{0,0}^{\left(4\right)}  \nonumber
\, .
\end{eqnarray}

\noindent
The hopping operators in Eq.(\ref {righthand}) mean that a vehicle can jump forward only if there are no vehicles at its right. In this
case, vehicles from outside have the priority on vehicles into the roundabout. As the ``both vehicles tossing the coin"
argument is not valid, we expect, on the basis of the  arguments presented in Fig.~\ref{steps}, that the right-hand rule
always gives a current across the junction lower than the wild setup. Indeed, we find that  this is the case. An indirect confirmation
comes from panels c) and d) of Fig.~\ref{merged}, where we do not consider the unrealistic wild case and compare
the ``right-hand" and ``roundabout" priority rules for $\beta=1$ and $\beta=0$, respectively. We observe that the
right-hand rule still survives as the best option in the small low-density corner of the phase diagram and that this
region increases with $\beta$. Also in this case, the results are in agreement with real-life traffic experience.

Before getting to the actual discussion of our results, let us make two remarks:
first, in the low-density regime, configurations like the one discussed in Fig.~\ref{steps} are
unlikely and, therefore, the roundabout effectiveness is reduced compared to the high-density regime, like in the previous case; second,
the ``double toss coin" argument cannot be applied to the right-hand rule, due the presence of the $\sigma_{0,0}$ operator in the
$L_{\rm hop}^{(\ell)}$ operator. Now, let us assume that at the same time we have a vehicle on one of the external sites,  say $\ell=4$,
and another vehicle on the internal site $\ell =3$. They both want to jump into $\ell=1$. If we allow the vehicle in $\ell=4$ to jump first
(right-hand rule), we are able to free the site and allow another vehicle to jump into it, thus increasing the current as vehicles can enter from both $\ell=4$ and $\ell=6$. On the other side,
letting the vehicle in $\ell=3$ jump first will not give the same benefit, as a vehicle from outside can jump into the
system only from $\ell=6$ that, in the low-density regime, is most likely empty, but not from $\ell=4$ that is occupied. At the end of the day, as far as the current
in the junction is concerned, the  presence of vehicles on dots $\ell=4,5,6$ is less convenient than the presence in dots
$\ell=1,2,3$, when few vehicles are around. This is not the case in the high-density
regime, when all the dots are expected to be occupied and the effect of setups like the one in Fig.~\ref{steps}
dominates, making the roundabout more effective. Clearly the ``polite'' roundabout, $\beta=1$, always wins against the
``no-turn signal'' roundabout, $\beta=0$. For this reason in panel d) of Fig.~\ref{merged} we compare the right-hand rule and the no-turn signal
roundabout, observing a similar behavior as in panel c) of Fig.~\ref{merged}, but with an enlarged (green)
region in which the right-hand priority rule produces a greater current.

\subsection{The priority intersection}
\label{priority}

The priority intersection is a quite common road configuration. We have a two-arm main road, that carries most of the traffic, and a
side road with a minor traffic flow. Consequently, we set $\Gamma^{(4)}=\Gamma^{(5)}=\Gamma_>$ and $\Gamma^{(6)}=\Gamma_<$. The other
coupling constants are chosen as $\gamma^{(\ell)}=\gamma$, $t^{(\ell)}=t$, with unbalanced $\mathcal{Z}_3$-broken exit-intention flux:
$\alpha^{(1)}=\alpha^{(2)}=0.75$, $\alpha^{(3)}=0.5$, being
$\Gamma ^{(\ell)}_1=\alpha ^{(\ell)}t$, $\Gamma ^{(\ell)}_2=(1-\alpha ^{(\ell)})t$.
In the priority intersection, vehicles from sites $\ell=4,5$ have the priority over vehicles from $\ell=6$ and vehicles from $\ell=4$
that want to turn left have to give the priority to vehicles from $\ell=5$. In our model this is realized starting from the wild rules
defined in the first part of this Section and by making the replacements:

\begin{eqnarray}
L_{\rm hop}^{\left(1\right)} & \rightarrow & \sqrt{t^{(1)}}\sigma_{1,0}^{\left(2\right)}\sigma_{0,2}^{\left(1\right)}\sigma_{0,0}^{\left(5\right)}  \nonumber \\
L_{\rm hop}^{\left(3\right)} & \rightarrow & \sqrt{t^{(3)}}\sigma_{1,0}^{\left(1\right)}\sigma_{0,2}^{\left(3\right)}\sigma_{0,0}^{\left(4\right)}  \\
L_{\rm hop,j}^{\left(6\right)} & \rightarrow & \sqrt{\Gamma_j^{(3)}}\sigma_{j,0}^{\left(3\right)}\sigma_{0,1}^{\left(6\right)}
\left( \sigma_{0,0}^{\left(2\right)}+\beta \sigma_{1,1}^{\left(2\right)} \right)\nonumber \, ,
\end{eqnarray}

\begin{figure}
\center
\includegraphics*[width=0.4 \linewidth]{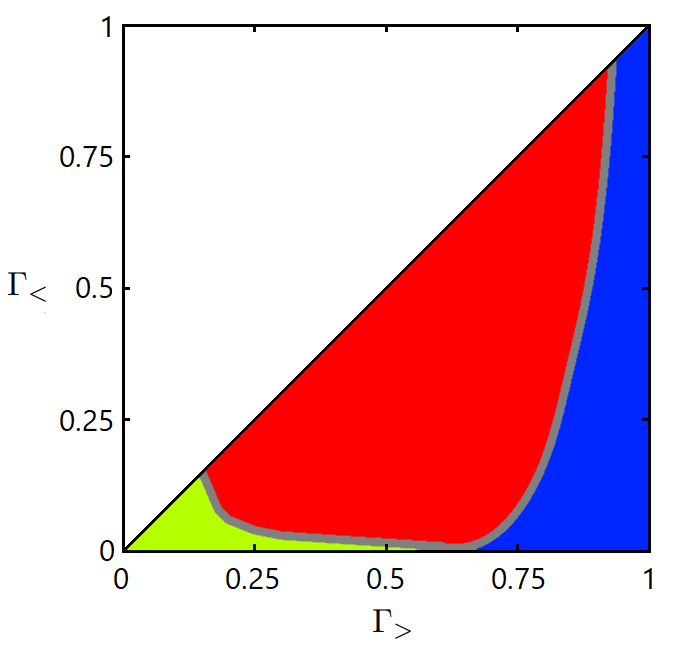}
\caption{Phase diagram of a priority rule intersection with two major priority fluxes, $\Gamma^{(4)}=\Gamma^{(5)}=\Gamma_>$, and a minor secondary flux,
$\Gamma^{(6)}=\Gamma_<$. At each point we plot which priority rule, between the priority intersection one (yellow region), the perfect roundabout
(red region) (both with $\beta =1$) and the traffic-light (blue region), exhibits the higher current across the system. We assume that the main
roads ($\ell=4,5$) accommodate most of the flow, setting $\alpha^{(1)}=\alpha^{(2)}=0.75$. The flux through the secondary road is taken symmetric,
setting $\alpha^{(3)}=0.5$. The other parameters are $t^{(\ell)}=1$, $\Gamma_{\rm C}=0.1$, $\Gamma_j^{(\ell)}=1$, $\gamma^{(\ell)}=3$.
For the traffic-light case, we set $T=2$ and $t_1=0.75T$.}
\label{priority2}
\end{figure}

\noindent
with $\beta $ measuring the promptness of the drivers to announce their direction (as in the roundabout case, $\beta =1$ means that drivers correctly use turn signals).
In Fig.~\ref{priority2} we compare the priority intersection rule with $\beta =1$ with the current that we would have by replacing
the junction with  the ``polite'' ($\beta=1$) roundabout or by setting a traffic light that regulates the major and minor fluxes.
In fact, within our model the traffic light is easily introduced by setting time-dependent coupling constants that are alternatively zero,
that is $\Gamma^{(4)}=\Gamma^{(5)}=0$ if $t\in\left[nT,nT+t_{1}\right]$ and $\Gamma^{(6)}=0$ if $t\in\left[nT+t_{1},(n+1)T\right]$, and then computing
the mean current over the periodicity of the traffic light $T$.
In doing this, we  assume that both $t_1$ and $T$ are much larger than the characteristic time in which the current reaches its stationary value.
We observe that priority intersection rules are more efficient at low density, {\it i.e.} for small $\Gamma _{>}$ and $\Gamma _{<}$ but,
at high-density, the roundabout and traffic light compete as the best configuration, depending on the values of $\Gamma_>$ and $\Gamma_<$.
This result shows a remarkable agreement with real-life traffic data and classical traffic models, as one can for instance verify by comparing
our Fig.~\ref{priority2} with Figs.~8-11 of Ref.~\cite{class_round_3}.

\section{Conclusions}
\label{conclusion}

We proposed to model a three-road junction  by means of a graph of multilevel quantum dots coupled to external
reservoirs. Hopping between quantum dots represent the flux of vehicles inside the junction. Jumps from/to external reservoir to/from internal dots
represent vehicles entering/exiting the junction. Each level of the quantum dots represents a possible kind of vehicle and/or destination. The vehicle
destination is chosen from the beginning, being an inner property of the vehicle, giving rise to complex fluxes combinations that are not observed in a junction
of (two-level) spin or fermionic quantum chains, where a particle can stochastically be annihilated from any possible exit point.
We first studied a three-dot model, which represents vehicles entering and exiting a junction without any rule.
Even if it looks like a minimal setup, such model already exhibits the main features of  a typical traffic-flow diagram, that is the existence of a critical
density of vehicles beyond which a traffic-jam phase occurs and the current (or flux of vehicles) decreases (Fig.~\ref{foundamental}).
In addition, in analyzing the current through the intersection as a function of the incoming flux, $\Gamma$,  we found an optimal working point, in which the current
(or the traffic-flow) is maximal (Fig.~\ref{current-density}).

However, to better describe a realistic system,  we added to the minimal model above three external dots, which
allow to mimic various priority rules used in road junctions. In this way, we
have been able to implement
a roundabout, a right-hand priority junction and an intersection between a major and a minor road by means of
 a six-dot model, with three levels for each dot.
For this model we could report in various cases (shown in panels a), b), c), and d) of Fig.~\ref{merged} and in Fig.~\ref{priority2}) which configuration shows the greater current
(or, in other words,   is more effective in letting vehicles  cross the junction) depending on incoming fluxes of vehicles from the external roads.
Our results qualitatively agree with real data, as reported for instance in Ref.~\cite{class_round_3}.

An advantage of the model we proposed is that it provides a microscopic simulation (by means of a system of quantum dots) of a macroscopic system, a junction
between three roads. It would be interesting to compare other predictions of our model or generalizations of it to real data of traffic in junctions.
The aim is to investigate the use of space/time correlations to predict traffic congestions as our approach allows for exactly computing $N$-time correlators of nonlocal operators~\cite{forecast_2}.
If these further comparisons, as the ones we made in this paper, will be successful, one could propose dot models
(and possibly, a realization of them in a laboratory) as microscopic simulators of real traffic situations.

Further extensions of our work are related to the possibility to consider more complex graphs, as for instance
four-leg intersections, car-pedestrian intersections (in terms of stochastic time-dependent coupling strengths), interaction between different kinds of vehicles.
In addition one could investigate the light-signalized intersection, trying to predict the optimum value for the time dependence of the hopping parameters.

Finally, within the Lindblad master equation formalism, it is possible to introduce a Liouvillian quantum term, describing quantum coherent evolution
of the density matrix. This would allow to investigate coherent and stochastic evolution on the same footing, exploring the crossover between quantum and classical steady states.

\paragraph{Acknowledgements}
We thank P. Pantano for insightful comments on the manuscript and C. Paletta for useful discussions at the early stage of this work.

\paragraph{Funding information}
A. N. was financially supported  by POR Calabria FESR-FSE 2014/2020 - Linea B) Azione 10.5.12,  grant no.~A.5.1.
A. N., D. G. and M. R. acknowledge  financial support  from Italy's MIUR  PRIN projects TOP-SPIN (Grant No. PRIN 20177SL7HC).

\bibliography{jam_biblio_2}

\nolinenumbers
\end{document}